\documentclass[useAMS,usenatbib]{mn2e}
\usepackage{times}
\usepackage{graphicx}
\usepackage{rotating}

\title[Mode identification for Balloon\,090100001]
{Mode identification for Balloon\,090100001 using combined multicolour photometry and spectroscopy}
\author[A.\,Baran, A.\,Pigulski, S.J.\,O'Toole]
{A.\,Baran$^{1,2}$\thanks{E-mail: andy@astro.as.ap.krakow.pl}, A.\,Pigulski$^{3}$, S.J.\,O'Toole$^{4}$ \\
$^{1}$Cracow Pedagogical University, ul.\,Podchor\c{a}\.zych\,2, 30\,--\,084 Krak\'ow, Poland \\
$^{2}$Toru\'n Centre for Astronomy, ul.\,Gagarina 11, Toru\'n, Poland \\
$^{3}$Instytut Astronomiczny Uniwersytetu Wroc{\l}awskiego, ul.\,Kopernika 11, 51\,--\,622 Wroc{\l}aw, Poland \\
$^{4}$Anglo-Australian Observatory, PO Box 296, Epping 1710, Australia}
\begin{document}

\date{Accepted . Received ; in original form }

\pagerange{\pageref{firstpage}--\pageref{lastpage}} \pubyear{2007}

\maketitle

\label{firstpage}

\begin{abstract}
In this paper, we show that method of mode identification using combined multicolour photometry
and spectroscopy can be successfully applied to the pulsating subdwarf B star Balloon\,
090100001. The method constrains the spherical degree, $\ell$.We confirm that the dominant mode
is radial and we show that for some other modes the method provides values of $\ell$ consistent with
the observed rotationally split triplet. Moreover, we derive a radius variation of 1.7 per cent
for the dominant mode. The identification opens the possibility for constraining the internal
structure of the star by means of seismic methods.
\end{abstract}


\section{Introduction}
About a decade ago, pulsations in hot subdwarfs (sdB) were discovered by 
astronomers at the South African Astronomical Observatory \citep[][and subsequent papers in the series]{kilkenny97}. At present, 
there are two classes of pulsating hot subdwarfs (sdBVs) known: V361\,Hya and PG\,1716 stars. 
The former pulsate in p modes which have short periods,
typically a few minutes long, the latter, in g modes which have periods an order of magnitude 
longer. Some V361\,Hya stars show relatively large amplitudes, reaching or even exceeding
50~mmag in $V$, while for PG\,1716 stars only low-amplitude modes are observed. The two 
classes also have slightly different effective temperatures, V361~Hya
stars are on average
hotter than PG\,1716 stars.

Considering the poorly understood evolutionary history of sdB stars,
it is very important to obtain information on their internal
structure whenever possible. The discovery of pulsations
in sdB stars has opened a new avenue to probe their interiors by means of seismic 
techniques. However, a prerequisite of a successful application of asteroseismology is 
detection of at least several modes which are properly identified in
terms of their pulsation geometry. The identification can be carried out using different methods that 
utilize photometric or spectroscopic data or a combination of both.

For sdBV stars, spectroscopic methods have an extra layer of
difficulty because the currently known objects are fainter than
$V\,\approx$\,12~mag, and they 
all have very short pulsation periods. Good time-resolved
spectroscopic observations can therefore only be obtained using large telescopes. 
For this reason, there have been many attempts to use only (multicolour) photometric data to identify modes
in sdBV stars. In particular, methods that compare observed and theoretical behaviours
of pulsation amplitude with wavelength have been used to constrain the spherical degree, $\ell$.
For example, such applications have been presented by \cite{koen98} for KPD\,2109+4401, 
\cite{jeffery04} for HS\,0039+4302 and \cite{jeffery05} for
PG\,0014+067. Unfortunately, photometric data are limited because the
amplitude ratios of modes with different $\ell$ sometimes have 
very similar wavelength dependence \citep*[see][]{ramach04},
and therefore this does not lead to an adequate discrimination of $\ell$. In the best case, 
only some possibilities could be rejected \cite[see Sect.~5 or][]{tremblay06}. 

Despite the difficulties in obtaining good-quality, time-resolved spectra for sdBV stars,
spectroscopic time-series data have been obtained for some of the
brightest objects: for example, 
PG\,1605+072 = V\,338~Ser \citep*{ot00,ot02,ot03,ot05,woolf02,falter03}, KPD\,2109+4401 
and PB 8783 \citep{jeffery00}, PG1627+017 \citep{for06} or Balloon 090100001 \citep[hereafter Bal09;][hereafter TO06]{telting06}.
In this paper we show that the method
of \cite*{dasz03} can be successfully applied to the extremely interesting pulsating subdwarf Bal09.

\section{Balloon 090100001}
\begin{figure}
\includegraphics[width=83mm]{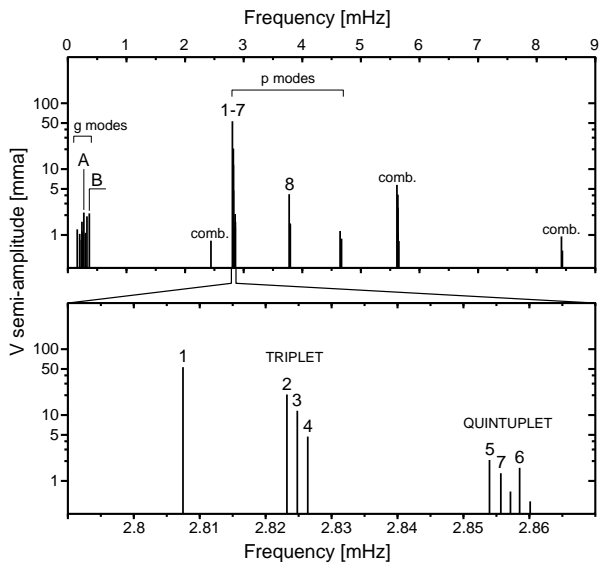}
\caption{\small A schematic Fourier spectrum of the 2004 $B$-filter data of Bal09. 
Note that semi-amplitudes are given on a logarithmic scale. Modes with the largest
amplitudes are labelled as in Paper I and Tables 1 and 2. Two components 
of the quintuplet (not labelled in the bottom panel) were detected only in the Fourier spectrum
of 2005 data.}
\label{ft}
\end{figure}

Bal09 is presently the brightest known sdBV star. 
Its pulsations were found by 
\cite{oreiro04} from a single night of photometric data. 
The main mode detected in this star has a semi-amplitude of about 
60~mmag in $B$ and a period of about 6 min, quite long for a V361\,Hya star.  
Thus, Bal09 became an obvious target for a more detailed study.  The follow-up photometry carried out in 
2004 \citep[hereafter Paper I; \citeauthor{oreiro05} \citeyear{oreiro05}]{baran05} revealed several interesting features in its 
frequency spectrum (Fig.~\ref{ft}). First of all, among about 20 detected modes, an equidistant 
triplet was found close to the main mode. If interpreted in terms of 
rotational splitting, it would imply a rotational period of about 7 d (Paper I).
Even more importantly, several low-frequency g modes were found. 
This made the star one of two presently known hybrid V361\,Hya/PG\,1716 stars; 
the other one is HS\,0702+6043 \citep{schuh06}.

In 2005, a large multisite campaign was carried out on Bal\,09 (Baran et al., in preparation). 
The new data allowed the detection of a quintuplet (see Fig.~\ref{ft}) whose
three strongest components had already been
detected in the 2004 data. In addition, it turned out that 
the amplitudes of some modes change. However, the most surprising discovery
was finding that splittings of multiplets changed 
considerably between 2004 and 2005 \citep{baran07}. As far as we are aware, this is the first 
clear evidence for such a change in any pulsating star. This finding
poses a problem for the interpretation of splittings only in terms of rotation.
However, we believe that rotation is the main factor that causes splittings, while another 
(presently unknown) effect(s) causes its small variations. We therefore refer to 
the multiplets as rotationally split.

\section{Observations and analysis}
The amplitude changes detected in Bal\,09 and other sdBV stars suggest
that if one wants to combine photometry and spectroscopy for mode
identification, the data must be obtained simultaneously.
For this reason, we decided to use the multiband photometry of Paper I
and spectroscopy of TO06 which were
coincidentally acquired at the same time, during 2004 August and September.

The $UBVR$ photometry of Paper I was carried out at two sites, Mt.~Suhora
and Loiano Observatories. About 120~h of data were gathered
through each filter. The observations were spread over about 40 d,
resulting in a frequency resolution of better than 0.5~$\mu$Hz. 
Details of the Fourier analysis of these data can be found 
in the Paper I. As found in that paper, the frequency spectrum
of Bal\,09 shows three regions of p modes (around 2.8, 3.8 and 4.7 mHz),
a group of low-frequency (below 0.5 mHz) g modes and three regions
where combination modes occur, labeled `comb.' in Fig.~\ref{ft}.

\begin{figure*}
\includegraphics[angle=90]{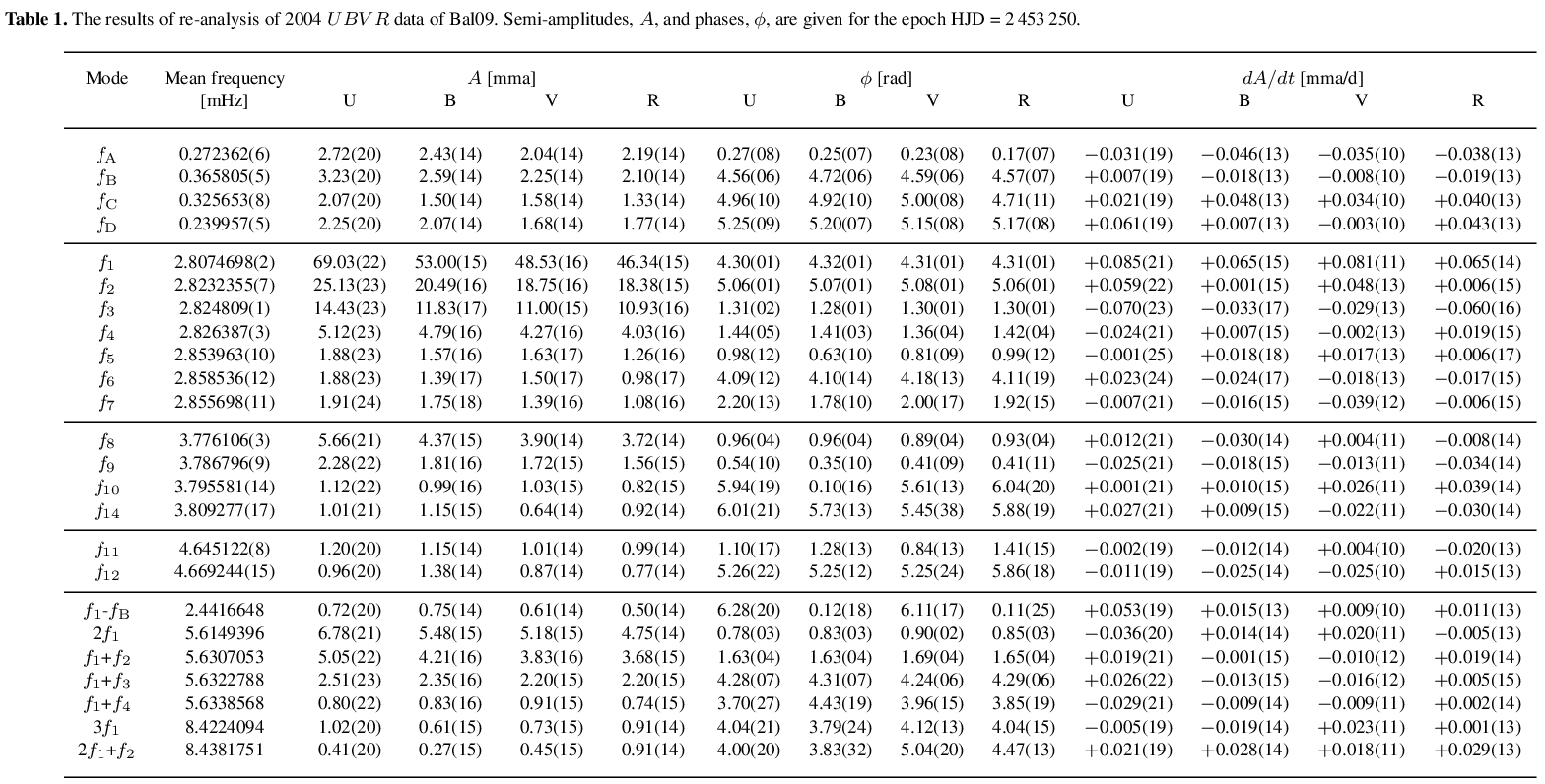}
\label{res_phot}
\end{figure*}

Because the changes in amplitude appeared to be more pronounced in the 2005 campaign data than we
found in Paper I, we decided to reanalyze
the 2004 photometry. This time, however, we allowed for
linear amplitude changes. In addition, the outliers were removed with
a slightly more stringent criterion than in Paper I. This resulted in a lower detection
threshold. We also decided to accept only those modes which were
detected in all four filters. As a consequence, the list of modes
we present in Tab.~\ref{res_phot} is shorter than in Paper I.

As mentioned above, spectroscopic observations of Bal\,09 were
carried out by TO06, simultaneously with the 2004 photometry. During seven nights spanning over 38 d, they obtained over 
2500 low-resolution spectra. Importantly the cycle time 
was only 43~s long, allowing good sampling of the radial velocity curve.
Details of the reduction and analysis can be found in TO06. 
We have reanalyzed the radial-velocity data of TO06 in order to obtain 
the fit parameters in a manner consistent with the photometry. As the 2004 photometric data suffered from 
aliasing much less than the radial-velocity data of TO06, we always adopted a frequency measured from photometry, even
if the alias peak was found to be higher than that at the ``real'' frequency. In total, we detected 
14 modes in radial velocity; their parameters are presented in
Tab.~\ref{res_spec}. We found the same $p$ modes as TO06: $f_1$, the triplet ($f_2$, $f_3$, and $f_4$), $f_8$
and two combination frequencies, 2$f_1$ and $f_1 + f_2$. In the low-frequency domain, we also 
detected seven modes above the detection level. Unfortunately, none of these could be unambiguously
identified with the modes detected in photometry, and only some of these were the same as those reported by TO06.
Apparently, the rich spectrum of $g$ modes and their very low amplitudes make the aliasing a very 
severe problem during the prewhitening procedure.

\begin{table}
\centering
\caption{Results of the sine-curve fit to the low-resolution spectroscopic 
data of TO06. As for the photometric data (Tab.~\ref{res_phot}),
the phases are given for the epoch HJD = 2\,453\,250. $\Delta\phi$ = $\phi_{\rm RV} - \phi_{\rm phot}$ 
denotes difference between radial-velocity and photometric (averaged over all four filters) phases.}
\begin{tabular}{ccccc}
\hline\noalign{\smallskip}
Mode & Frequency & Semi-amplitude & $\phi_{\rm RV}$ & $\Delta\phi$ \\
    &   [mHz]   &   [km\,s$^{-1}$]  & [rad] &      [rad]     \\
\noalign{\smallskip}\hline\noalign{\smallskip}
$f_{\rm 1}$ & 2.8074698 & 19.16(20) & 5.74(01) & $+$1.43(02) \\
$f_{\rm 2}$ & 2.8232355 &  5.97(25) & 0.33(04) & $+$1.54(04) \\
$f_{\rm 3}$ & 2.824809  &  3.64(26) & 2.66(07) & $+$1.36(07) \\
$f_{\rm 4}$ & 2.826387  &  1.39(24) & 3.12(18) & $+$1.71(18) \\
$f_{\rm 8}$ & 3.776106  &  1.11(18) & 2.38(17) & $+$1.45(17) \\
2\,$f_{\rm 1}$          & 5.6149396 & 1.23(19) & 1.98(15) & $+$1.14(15) \\
$f_{\rm 1}$+$f_{\rm 2}$ & 5.6307053 & 1.10(19) & 3.28(17) & $+$1.63(17) \\
\hline
TO06       & 0.32812(3) & 2.12(19) &  5.37(09) & --- \\
TO06 alias & 0.36614(4) & 1.15(19) &  3.27(17) & --- \\
           & 0.16944(3) & 1.36(19) &  0.63(14) & --- \\
TO06       & 0.46406(3) & 1.16(19) &  4.46(16) & --- \\
           & 0.29727(4) & 1.51(20) &  3.64(13) & --- \\
           & 0.26247(4) & 1.21(19) &  1.81(16) & --- \\
\noalign{\smallskip}\hline
\end{tabular}
\label{res_spec}
\end{table}

However, three g modes known from photometry, $f_{\rm A}$, $f_{\rm B}$ and $f_{\rm C}$ according 
to the designations used in Tab.~\ref{res_phot}, were found by TO06 in the variations of the 
equivalent widths (EW) of the Balmer H$\gamma$--H9 lines.
The reason why the same modes were not detected in radial velocities can be qualitatively explained if we recall
that for g modes the horizontal motions dominate the radial motions, while for p modes the opposite is true. 
As all have small amplitudes, it is therefore quite possible that different modes are seen in photometry
(and EW changes) and radial velocities. As shown by \cite{for06}, because of the small amplitudes of g modes, 
it is very difficult to detect them in radial-velocity data.
For modes observed in both photometry and spectroscopy, simultaneous
data allow us to derive phase lags between 
light and radial-velocity variations. This parameter is related to the non-adiabacity 
of pulsations. The phase lags for modes detected in photometry and spectroscopy
are given in Tab.~\ref{res_spec}. The weighted mean of the phase lag for five independent $p$ modes is equal
to 1.45 $\pm$ 0.03~rad or 83 $\pm$ 2$^{\circ}$. For the dominant mode, $f_1$,
it is very similar, 1.43 $\pm$ 0.02~rad. Because this mode is radial (see Sect.~5), the phase of maximum 
brightness corresponds roughly to the phase of minimum radius.
In other words, we can conclude that in V361\,Hya stars the brightness variations are
dominated by the temperature effects. Adopting projection factor equal to 1.4 as
typical for an sdB star \cite[see][]{moroje01} and a radius of 0.25\,$R_\odot$,
we estimate the full range of radius variations for the dominant mode
in Bal09 to be about 1.7 per cent.

\section{The method of mode identification}
Having derived frequencies, amplitudes and phases we can now apply the
method (described in detail by \cite{dasz03}) to 
discriminate $\ell$ parameters for the modes detected in our
photometry and spectroscopy of Bal09. 

In the simplest form, the method of \cite{dasz03} utilizes a set of linear equations, 
of the form
\begin{equation}
\mathcal{D}_{\lambda,\ell}(\tilde{\varepsilon} f) + \mathcal{E}_{\lambda,\ell} 
\tilde{\varepsilon} = A_\lambda
\label{eq1}
\end{equation}
to derive two complex parameters, $\tilde{\varepsilon}$ and $f$. The coefficients
$A_\lambda$ are taken from observations; these are amplitudes, $a_\lambda$, and 
phases, $\phi_\lambda$, written in a complex form, $A_\lambda = a_\lambda 
\exp(i\phi_\lambda)$, where $\lambda$ denotes a given photometric band.
However, $\mathcal{D}_{\lambda,\ell}$ and $\mathcal{E}_{\lambda,\ell}$ 
depend on the the flux and limb darkening derivatives over effective temperature
and surface gravity, and have to be calculated using model stellar atmospheres. In principle,
observations in two photometric bands are sufficient to solve the set of equation (1)
once the spherical degree $\ell$ is assumed. However, if observations in more than 
two photometric bands are available, the set becomes overconditioned and can 
be solved by means of the least-squares method. Usually, however, we are not interested in
the values of $\tilde{\varepsilon}$ and $f$, but in discriminating $\ell$. In 
order to achieve this, the equations are solved for different values of $\ell$ and then
a goodness-of-fit statistic like reduced $\chi^2$ is used to decide on discrimination.

Unfortunately, as was shown by \citet*[][see also Sect.~5]{dasz05}, using merely	
photometric data rarely leads to unambiguous discrimination of $\ell$. It is much better
to combine photometric and spectroscopic data adding another equation of set, that is
\begin{equation}
\mathcal{F}_{\lambda,\ell} \tilde{\varepsilon} = \mathcal{M}_{1,\lambda}
\end{equation}
In Eequation (2), $\mathcal{M}_{1,\lambda}$ represents the first moment of spectral
line variations, (i.e. radial velocities), while $\mathcal{F}_{\lambda,\ell}$ is again
calculated from the model atmospheres. The solution of a set of equations (1) and (2) 
leads in this case to a much better discrimination of $\ell$ (see Section~5).

The model-dependent coefficients in equations (1) \& (2) and, consequently,
the discrimination of $\ell$ depend on the model parameters, in
particular effective temperature, surface gravity and metallicity. For
Bal09, the effective temperature and surface gravity were derived by
\cite{oreiro04}, T$_{\rm eff}$ = 29\,450 $\pm$ 500~K, $\log g$ = 5.33
$\pm$ 0.10~dex and \cite{telt06}, T$_{\rm eff}$ = 28\,700 K, $\log g$ = 5.39~dex. 
We therefore calculated flux distributions in the
range between 300 and 1000~nm for a grid of models with T$_{\rm eff}$
ranging from 28\,100 and 31\,000~K and $\log g$ from 5.2 to
5.5~dex. To do this, a grid of fully metal line-blanketed local thermodynamic 
equilibrium (LTE) stellar atmosphere models was calculated with the ATLAS9
code of \citet{kurucz92}. Flux distributions
were then calculated using Michael Lemke's version of the LINFOR
program (originally developed by Holweger, Steffen, and Steenbock at
Kiel University), similar to the synthetic spectra used by
\citet{othe06}. The Kurucz line list was used as a source of
oscillator strengths and damping constants for all metal lines. Only
lines that have been observed experimentally were included. The
resulting flux distributions were multiplied by transmission functions
of the filters in order to obtain integrated fluxes in the four
photometric bands, $UBVR$, that we used.                                   

For the same grid of models, we also calculated the specific intensities as a function 
of $\mu$ = $\cos\theta$, where $\theta$ was the angle between the line of sight 
and the normal to the stellar surface in a given point of the stellar
disc. Using these intensities, the limb darkening law was derived,
fitting the coefficients $c$ and $d$ of a function
\begin{equation}
I(\mu)\,=\,I(1)[1-c(1-\mu)-d(1-\sqrt{\mu})]
\label{eq2}
\end{equation}
\citep{dicg92} which was found to reproduce best the calculated changes 
of $I(\mu)$. This was carried out for different combinations of input 
parameters of the model, thus allowing the calculation of neccessary
derivatives.

The flux distribution calculated from a model atmosphere depends also
on helium abundance, metallicity and microturbulent velocity. 
Following the determination of \cite{oreiro04}, helium abundance was assumed
to be $\log$[n(He)/n(H)] = $-$2.54. In addition, we assumed solar metallicity and
microturbulent velocity $\xi$ = 0~km\,s$^{-1}$. The latter two
parameters have not been determined yet for Bal09, but are typical
values for an sdB star \citep[e.g.][]{othe06}.

\section{Discrimination of the $\ell$ parameter}
\begin{figure}
\includegraphics[width=83mm]{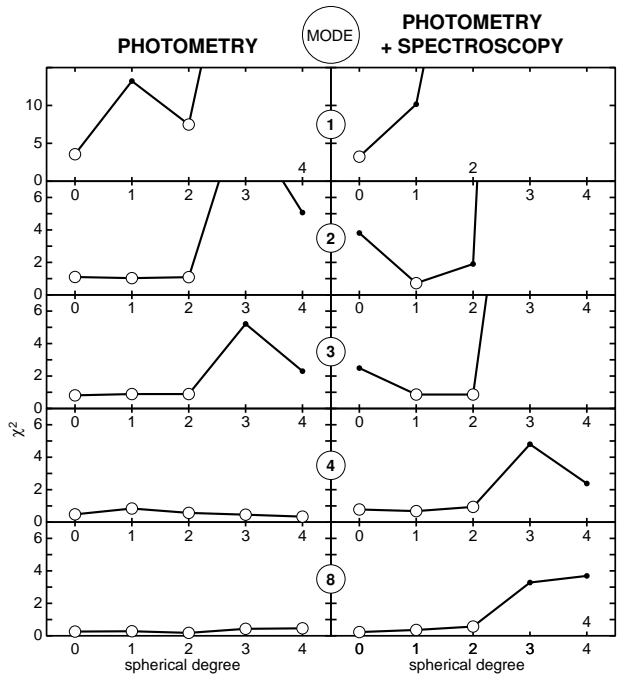}
\caption{\small Reduced $\chi^{\rm 2}$ as a function of the spherical degree $\ell$ for 2004
data of Bal09 and five modes detected both in photometry and spectroscopy. 
The modes are labelled with the encircled designation in the middle 
(see~Tables \ref{res_phot} and \ref{res_spec}). Left panels show
the results when only photometric data were used in calculations, and right panels when
photometric and spectroscopic data were combined. The $\ell$ value(s) that we regard as
possible for a given mode are shown with large circles.}
\label{chi2_ubvr+spektr}
\end{figure}

As a first step, we used the method presented in Section~4 in 
its simplest form (i.e. taking into account only photometric amplitudes 
and phases in four bands; see Tab.~\ref{res_phot}).  
The results for six modes detected both in photometry and radial velocities
are presented in Tab.~\ref{chi2_tab_ubvr} and in the left panels of
Fig.~\ref{chi2_ubvr+spektr}. We limited our calculations to $\ell \leq$ 4 because,
as a result of averaging over the stellar disc, modes with larger $\ell$ are not expected
to be visible in photometry.

\begin{table}
\centering
\caption{The results of the application of the method of \citet{dasz03} for Bal09 2004 
photometric data: reduced $\chi^{\rm 2}$ values for five modes with the 
largest amplitudes as a function of the spherical degree $\ell$.}
\begin{tabular}{crrrrrr}
\hline\noalign{\smallskip}
$\ell$ & $f_{\rm 1}$ & $f_{\rm 2}$ & $f_{\rm 3}$ & $f_{\rm 4}$ & $f_{\rm 8}$\\
\noalign{\smallskip}\hline\noalign{\smallskip}
 0 &  3.56 &  1.10 & 0.81 & 0.48 & 0.26 \\
 1 & 13.22 &  1.03 & 0.89 & 0.84 & 0.28 \\
 2 &  7.48 &  1.09 & 0.89 & 0.57 & 0.18 \\
 3 & 42.12 & 10.67 & 5.21 & 0.46 & 0.43 \\
 4 & 28.39 &  5.07 & 2.30 & 0.34 & 0.46 \\
\noalign{\smallskip}\hline
\end{tabular}
\label{chi2_tab_ubvr}
\end{table}

As can be seen from Fig.~\ref{chi2_ubvr+spektr} and Tab.~\ref{chi2_tab_ubvr},
there is only little discrimination of $\ell$ for $f_1$, $f_2$ and
$f_3$ and none for $f_4$ and $f_5$. For $f_1$, $\ell$ = 0 or 2 are
possible, while for $f_2$ and $f_3$, $\ell >$ 2 are excluded.

The poor discrimination of $\ell$ when using photometric data
can be understood when considering the amplitude
ratios versus wavelength calculations carried out by \citet{ramach04}. In
addition, three of the four photometric 
bands we used are located in the Paschen continuum. The resulting 
equations for different bands might therefore be highly correlated, which 
will not lead to a good discrimination of $\ell$.

It also has to be pointed out that the reduced $\chi^2$ statistics we use to discriminate $\ell$,
does not have, in our case, its proper characteristics, as the set of equation (1) is
not well overconditioned; it has only four degrees of freedom. 
For this reason, we did not
set a given threshold in $\chi^2$ to indicate possible values of $\ell$. Instead, we
do this arbitrarily, comparing the values of $\chi^2$ for different values of $\ell$; those
with the smallest $\chi^2$ are regarded as possible (see Fig.~\ref{chi2_ubvr+spektr}). 

In the next step, we added the radial velocity data as an additional constraint via equation~(2) 
and repeated the procedure. The results are shown in the right panels of 
Fig.~\ref{chi2_ubvr+spektr} and in Tab.~\ref{chi2_tab_ubvr+spektr}.
As equation~(2) contains only $\tilde{\varepsilon}$ as a free parameter, it is expected that 
$f$ and $\tilde{\varepsilon}$ will be much better constrained than previously and, consequently,
$\ell$ will be much better discriminated. As can be seen from Fig.~\ref{chi2_ubvr+spektr}, 
this is indeed the case.
Adding spectroscopy improves the situation in the sense that observational equations become
less correlated, which allows reliable estimation of $\varepsilon$ and $f$, but not as far as the
$\chi^2$ statistics is concerned. For combined data, the problem has two more (i.e. six) degrees
of freedom, still too small a number to say the problem is well defined. Because of that, we would
like to avoid giving a level of significance, as it would be irrelevant. Nevertheless,
relative numbers tell us which values of $\ell$ are acceptable, although, as we have pointed out, our
choice is arbitrary.

\begin{table}
\centering
\caption{The same as in Tab.~\ref{chi2_tab_ubvr}, but for the combined photometric 
and spectroscopic data.}
\begin{tabular}{crrrrr}
\hline\noalign{\smallskip}
$\ell$ & $f_{\rm 1}$ & $f_{\rm 2}$ & $f_{\rm 3}$ & $f_{\rm 4}$ & $f_{\rm 8}$ \\
\noalign{\smallskip}\hline\noalign{\smallskip}
 0 &   3.88 &  4.44 &  3.03 & 0.96 & 0.22 \\
 1 &   9.03 &  0.80 &  0.94 & 0.65 & 0.35 \\
 2 &  47.73 &  1.99 &  1.06 & 0.51 & 0.55 \\
 3 & 218.07 & 77.42 & 28.11 & 4.63 & 3.56 \\
 4 &1526.90 &101.09 & 35.45 & 6.03 & 6.03 \\
\noalign{\smallskip}\hline
\end{tabular}
\label{chi2_tab_ubvr+spektr}
\end{table}

The best and least ambiguous discrimination was obtained for the 
dominant mode, which appears to be radial ($\ell$ = 0). 
For the strongest component of the triplet, $f_2$, $\ell$ = 1 
is definitely the best solution, exactly as expected. 
For the two remaining triplet components, $f_3$ and $f_4$, 
$\ell$ is not unambiguously identified, but $\ell$ = 1 
is one of the possibilities. Therefore, we conclude that, 
as already suggested in Paper I, the main mode in Bal09 
is radial, while the triplet represents a rotationally 
split $\ell$ = 1 mode.

Table \ref{chi2_all} summarizes the results of our discrimination. 
In addition, we compare these results with those presented by \cite{tremblay06},
who used only photometric data to identify $\ell$ for the nine modes 
published in Paper I. They did not obtain a 
unique identification for any of these modes. For the best cases, they present two possible values
of $\ell$, similar to the results we obtained using only the
photometric data.
\begin{table}
\centering
\caption{The possible values of $\ell$ for five strongest modes in Bal09 
derived in this paper and by \citet{tremblay06}.}
\begin{tabular}{cccc}
\hline\noalign{\smallskip}
       Mode & This paper    & This paper     & \cite{tremblay06} \\
            & (photometry)  & (photometry +  &                   \\
            &               & spectroscopy)  &                   \\
\noalign{\smallskip}\hline\noalign{\smallskip}
$f_{\rm 1}$ & 0,\,2             & 0         & 0,\,1                 \\
$f_{\rm 2}$ & 0,\,1,\,2         & 1         & 1,\,2                 \\
$f_{\rm 3}$ & 0,\,1,\,2         & 1,\,2     & 1,\,2                 \\
$f_{\rm 4}$ & 0,\,1,\,2,\,3,\,4 & 0,\,1,\,2 & 1,\,2,\,4             \\
$f_{\rm 8}$ & 0,\,1,\,2,\,3,\,4 & 0,\,1,\,2 & 0,\,1,\,2,\,3         \\
\noalign{\smallskip}\hline
\end{tabular}
\label{chi2_all}
\end{table}

\begin{figure}
\includegraphics[width=83mm]{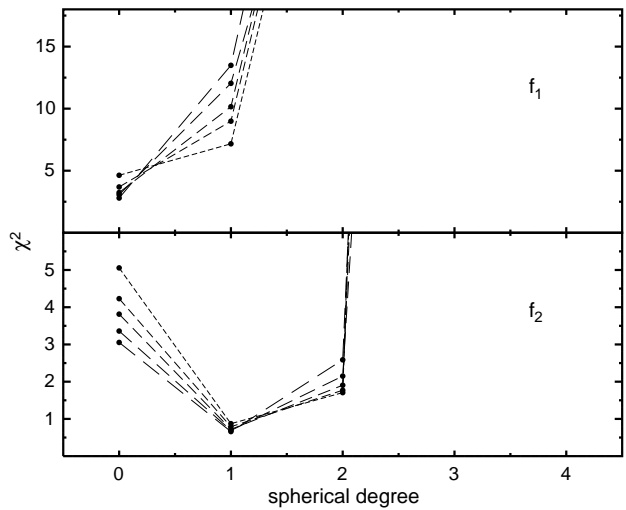}
\caption{\small Discrimination of the $\ell$ parameter for five different values of $T_{\rm eff}$,
combined photometry and spectroscopy and two strongest modes, $f_1$ (top) and $f_2$ (bottom).
Going from the shortest to the longest dashes results are shown for effective temperatures between 28\,500 and
30\,500~K in 500-K steps.}
\label{chi2_tests}
\end{figure}

We also checked how the discrimination we obtain depends on the parameters of the model. The result for
adopting five different effective temperatures from the range between 28\,500 and 30\,500~K is 
shown in Fig.~\ref{chi2_tests}. The change with $T_{\rm eff}$ does not greatly affect the
result of discrimination, although lower effective temperatures $\ell$ = 1 for $f_1$ and
$\ell$ = 2 for $f_2$ become alternatives to those given in Tab.~\ref{chi2_all}. 
A negligible effect was obtained while changing the mass between 0.47 
and 0.53~$M_\odot$ and radius between 0.23 and 0.26~$R_\odot$.

\section{Summary and conclusions}
Using photometric and spectroscopic data for Bal09, we have shown that 
the method of mode identification developed by \cite{dasz03} can be successfully applied 
to Bal09. We believe that, if simultaneous multicolour photometry and 
spectroscopy are carried out, the method will also allow mode identification in other sdBV stars. In Bal09, 
the spherical degree $\ell$ was unambiguously identified for the two modes with the
largest amplitudes. For several other modes, we limited the number of possible
values of $\ell$ to two or three. If we accept that the two observed 
multiplets are caused by rotational splitting, then the number of identified modes 
increases to nine: the triplet components have $\ell$ = 1, the quintuplet components, 
$\ell$ = 2. 

It is clear from our considerations that photometry itself 
is not sufficient to identify mode satisfactorily for sdBV stars, 
and that adding spectroscopic observations greatly improves the situation. For Bal09, it is certainly
worth carrying out another photometric and spectroscopic campaign
to identify modes with even lower amplitudes. However, our identification 
is probably sufficient to go a step further and attempt asteroseismology of this star.
However, for this purpose, an appropriate set of models is required. A
first attempt was presented in Paper I using the models of
\cite{charp02}. While we had no modes identified at that time,
the assumption that the main mode is radial now appears to be correct. 
With reliable identification
of modes from the 2.8~mHz group, a good seismic model that fits the identified
modes might be used to match (and identify) remaining modes. In order
to carry out a thorough asteroseismological analysis and potentially
learn new physics, it is important that more modes are identified
independent of pulsation models in sdBV stars. We have provided a
crucial piece of the puzzle, and the challenge is now for pulsation
theorists to find models that match both the pulsation
frequencies and mode identifications derived here.

\section*{Acknowledgments}
This project was partially supported by grant no. 1P03D 013 29 kindly provided 
by Polish MNiSW. Constructive criticism from the anonymous referee is also acknowledged.

\label{lastpage}


\begin{thebibliography}{}
\bibitem[\protect\citeauthoryear{Baran et al.}{2005}]{baran05} Baran A., Pigulski A., Kozie{\l} D., et al., 2005, MNRAS, 360, 737 (Paper I)
\bibitem[\protect\citeauthoryear{Baran et al.}{2007}]{baran07} Baran A., Oreiro R., Pigulski A., et al., 2007, 
Proc.~of the 15$^{\rm th}$ European White Dwarf Workshop EUROWD06, 372, 607
\bibitem[\protect\citeauthoryear{Charpinet et al.}{2002}]{charp02} Charpinet S., Fontaine G., Brassard P., Dorman B., 2002, ApJS, 140, 469
\bibitem[\protect\citeauthoryear{Daszy{\'n}ska-Daszkiewicz, Dziembowski \& Pamyatnykh}{Daszy{\'n}ska-Daszkiewicz et al.}{2003}]{dasz03} Daszy{\'n}ska-Daszkiewicz J., Dziembowski W.A., Pamyatnykh A.A., 2003, A\&A, 407, 999
\bibitem[\protect\citeauthoryear{Daszy{\'n}ska-Daszkiewicz, Dziembowski \& Pamyatnykh}{Daszy{\'n}ska-Daszkiewicz et al.}{2005}]{dasz05} Daszy{\'n}ska-Daszkiewicz J., Dziembowski W.A., Pamyatnykh A.A., 2005, A\&A, 441, 641
\bibitem[\protect\citeauthoryear{D\'{\i}az-Cordov\'es \& Gim\'enez}{1992}]{dicg92} D\'{\i}az-Cordov\'es, J., \& Gim\'enez, A., 1992, A\&A, 259, 227
\bibitem[\protect\citeauthoryear{Falter et al.}{2003}]{falter03} Falter S., Heber U., Dreizler S., et al., 2003, A\&A, 401, 289
\bibitem[\protect\citeauthoryear{For et al.}{2006}]{for06} For B.-Q., Green E.M., O'Donoghue D., et al., 2006, ApJ, 642, 1117
\bibitem[\protect\citeauthoryear{Jeffery \& Pollacco}{2000}]{jeffery00} Jeffery C.S., Pollacco D., 2000, MNRAS, 318, 974
\bibitem[\protect\citeauthoryear{Jeffery et al.}{2004}]{jeffery04} Jeffery C.S., Dhillon V.S., Marsh T.R. Ramachandran B., 2004, MNRAS, 352, 699
\bibitem[\protect\citeauthoryear{Jeffery et al.}{2005}]{jeffery05} Jeffery C.S., Aerts C., Dhillon V.S., Marsh T.R., G$\ddot{\rm a}$nsicke B., 2005, MNRAS, 318, 974
\bibitem[\protect\citeauthoryear{Kilkenny et al.}{1997}]{kilkenny97} Kilkenny D., Koen C., O'Donoghue D., et al., 1997, MNRAS, 285, 640
\bibitem[\protect\citeauthoryear{Koen}{1998}]{koen98} Koen C., 1998, MNRAS, 300, 567
\bibitem[\protect\citeauthoryear{Kurucz}{1992}]{kurucz92} Kurucz R.L.,
  1992, in: \textit{The Stellar Populations of Galaxies}, IAU
  Coll. 149, eds. B.Barbuy \& A. Renzini, Kluwer Academic Publishers,
  Dordrecht, p. 225.
\bibitem[\protect\citeauthoryear{Monta\~{n}\'es Rodriguez \& Jeffery}{2001}]{moroje01} Monta\~{n}\'es Rodriguez P., Jeffery C.S., 2001, A\&A, 375, 411
\bibitem[\protect\citeauthoryear{Oreiro et al.}{2004}]{oreiro04} Oreiro R., Ulla A., P\'erez Hern\'andez F., et al., 2004, A\&A, 418, 243
\bibitem[\protect\citeauthoryear{Oreiro et al.}{2005}]{oreiro05} Oreiro R., P\'erez Hern\'andez F., Ulla A., et al., 2005, A\&A, 438, 257
\bibitem[\protect\citeauthoryear{O'Toole \& Heber}{2006}]{othe06} O'Toole S.J., Heber U., 2006, A\&A, 452, 579
\bibitem[\protect\citeauthoryear{O'Toole et al.}{2000}]{ot00} O'Toole S.J., Bedding T.R., Kjeldsen H., et al., 2000, ApJ, 537, L53
\bibitem[\protect\citeauthoryear{O'Toole et al.}{2002}]{ot02} O'Toole S.J., Bedding T.R., Kjeldsen H., Dall T.H., Stello D., 2002, MNRAS, 334, 471
\bibitem[\protect\citeauthoryear{O'Toole et al.}{2003}]{ot03} O'Toole S.J., J{\o}rgensen M.S., Kjeldsen H., Bedding T.R., Dall T.H., Heber U., 2003, MNRAS, 340, 856
\bibitem[\protect\citeauthoryear{O'Toole et al.}{2005}]{ot05} O'Toole S.J., Heber U., Jeffery C.S., et al., 2005, A\&A, 440, 667
\bibitem[\protect\citeauthoryear{Ramachandran, Jeffery \& Townsend}{Ramachandran et al.}{2004}]{ramach04} Ramachandran B., Jeffery C.S., Townsend R.H.D., 2005, A\&A, 428, 209
\bibitem[\protect\citeauthoryear{Schuh et al.}{2006}]{schuh06} Schuh S., Huber J., Dreizler S., et al., 2006, A\&A, 445, 31
\bibitem[\protect\citeauthoryear{Telting \& {\O}stensen}{2006}]{telting06} Telting J.H., {\O}stensen R.H, 2006, A\&A, 450, 1149 (TO06)
\bibitem[\protect\citeauthoryear{Telting et al.}{2006}]{telt06} Telting J.H., {\O}stensen R.H., Heber U., Augusteijn T., 2006, Baltic Astron., 15, 235
\bibitem[\protect\citeauthoryear{Tremblay et al.}{2006}]{tremblay06} Tremblay P.E., Fontaine G., Brassard P., Bergeron P., Randall S.K., 2006, ApJS, 165, 551
\bibitem[\protect\citeauthoryear{Woolf, Jeffery \& Pollacco}{Woolf et al.}{2002}]{woolf02} Woolf V.M., Jeffery C.S., Pollacco D.L., 2002, MNRAS, 329, 497


\end{thebibliography}
\end{document}